# Emergence of intrinsic superconductivity below 1.178 K in the topologically non-trivial semimetal state of CaSn$_3$

Y.L. Zhu[1], J. Hu[1,2*], F.N. Womack[3], D. Graf[4], Y. Wang[1], P.W. Adams[3] and Z.Q. Mao[1*]

[1] Physics and Engineering Physics department, Tulane University, New Orleans, LA, 70118

[2] Department of Physics, University of Arkansas, Fayetteville, AR 72701

[3] Department of Physics and Astronomy, Louisiana State University, Baton Rouge, LA, 70803

[4] National High Magnetic Field Laboratory, Tallahassee, FL, 32310

Abstract

Topological materials which are also superconducting are of great current interest, since they may exhibit a non-trivial topologically-mediated superconducting phase. Although there have been many reports of pressure-tuned or chemical-doping-induced superconductivity in a variety of topological materials, there have been few examples of intrinsic, ambient pressure superconductivity in a topological system having a stoichiometric composition. Here, we report that the pure intermetallic CaSn$_3$ not only exhibits topological fermion properties, but also has a superconducting phase at ~1.178 K under ambient pressure. The topological fermion properties, including the nearly zero quasi-particle mass and the non-trivial Berry phase accumulated in cyclotron motions, were revealed from the de Haas–van Alphen (dHvA) quantum oscillation studies of this material. Although CaSn$_3$ was previously reported to be superconducting with $T_c$ =

4.2 K, our studies show that the $T_c$=4.2 K superconductivity is extrinsic and caused by Sn on the degraded surface, whereas its intrinsic bulk superconducting transition occurs at 1.178 K. These findings make CaSn$_3$ a promising candidate for exploring new exotic states arising from the interplay between non-trivial band topology and superconductivity, *e.g.* topological superconductivity.

* jinhu@uark.edu

*zmao@tulane.edu

# I. INTRODUCTION

Topological superconductivity (TSC) has attracted extensive interest due to its strong connection with Majorana fermions. Majorana fermions follow the Dirac equation and are particles which are their own antiparticles. The collective excitations on the surfaces of topological superconductors are believed to satisfy these conditions: the Dirac-like dispersion is guaranteed by the topological surface states and the particle-hole symmetry in the superconducting state results in indistinguishable electron and hole excitations. Therefore, the gapless surface states of a topological superconductor host Majorana fermions [1,2]. TSC can be realized via the superconducting proximity effect in a superconductor-topological insulator [3–6] or superconductor-semiconductor hybrid structure [7–9]. Signatures of Majorana fermions have indeed been reported in such heterostructures [9–13]. In particular, half-integer quantized conductance plateaus were recently observed in the quantum anomalous Hall insulator-superconductor hybrid system, suggesting one-dimensional chiral Majorana fermion modes [13].

In addition to the proximity effect-induced TSC, several other strategies have been utilized to achieve TSC in bulk materials. One is to use chemical doping in topological insulators. This has produced signatures of TSC in some systems, *e.g.* $Cu_xBi_2Se_3$ [14–16]. Point contact, which is generally used for tunneling measurements, has also been found to generate possible TSC in the topological semimetal $Cd_3As_2$ [17,18]. Another approach is to induce superconductivity in topological semimetals or insulators by applying high pressure [19–25]. However, it remains elusive whether such pressure-induced superconductivity is connected with TSC, since probing the Majorana surface modes of topological superconductors under pressure is a challenging task. In some cases, pressure is also likely to change the structure, thus resulting in trivial band topology [21]. These problems may be avoided by finding materials which display both non-trivial topological surface states and intrinsic superconductivity at ambient pressure. Some materials have been found to show such properties, such as (Y/Lu)PtBi [26,27], PbTaSe$_2$ [28–30], FeTe$_{0.55}$Se$_{0.45}$ [31], and β-PdBi$_2$ [32]. These materials not only have clear advantages for probing Majorana surface modes [30,31], but also offer opportunities to observe new superconducting phenomena, such as unconventional superconductivity with spin-3/2 pairing in YPtBi [33]. These exciting advances underscore the importance of searching for other systems having a non-trivial band topology in conjunction with a bulk superconducting phase.

In this article, we report on the discovery of a new topological semimetal showing intrinsic superconductivity at ambient pressure, *i.e.* CaSn$_3$. This material possesses the AuCu$_3$-type cubic crystal structure with the space group $Pm\bar{3}m$. Recent first-principles calculations predicted a topologically non-trivial electronic state for this material [34]; without considering spin-orbital coupling (SOC), its electronic band structure is found to host a 3D topological nodal line state with a unique drumhead like topological surface states protected by time reversal and mirror symmetries. When SOC is taken into account, each nodal line is predicted to evolve into two Weyl nodes [34]. This claim appears to conflict with a requirement for a Weyl state, i.e. either time reversal symmetry breaking or inversion symmetry breaking, since CaSn$_3$ does not satisfies either of these two requirements. Although this prediction needs to be further scrutinized, we have performed systematic quantum oscillation studies on this material through magnetic torque measurements, from which we indeed observed relativistic fermion behavior, suggesting

CaSn$_3$ should possess non-trivial band topology. Furthermore, we find that this material shows intrinsic bulk superconductivity around 1.178 K, while the previously reported non-bulk superconducting transition at 4.2K for this material [35] is extrinsic and caused by Sn present on its degraded surface. The presence of intrinsic superconductivity in a topologically non-trivial state makes CaSn$_3$ a promising new platform for the study of the interconnection between topological surface states and superconductivity.

## II. EXPERIMENTAL METHODS

CaSn$_3$ single crystals were synthesized using a self-flux method. The Ca pieces and Sn lumps were loaded into a baked Al$_2$O$_3$ crucible with the molar ratio of Ca/Sn = 1/4 and sealed in a quartz tube under high vacuum. The reagents were heated to 900 °C, kept at this temperature for 24 hours and then slowly cooled down to 400 °C at a rate of 4 °C/h. The single crystals can be obtained after removing the unreacted Sn flux through centrifugation; the inset of Fig. 1a shows an optical image of a typical CaSn$_3$ crystal. The cubic shape of the obtained crystals is consistent with the face-centered cubic structure for CaSn$_3$ (Fig. 1a) [35], which is further confirmed by X-ray diffraction measurements (XRD) (see supplementary Fig. S1). Moreover, we also found CaSn$_3$ single crystals are air sensitive, consistent with the previous report [35]. When a sample is left in air for more than 30 minutes, the surface color changes dramatically. XRD measurements suggest that the degraded surface layer contains Sn as shown in supplementary Fig. S1 and Fig. 1b in ref. [35], which causes the observation of the trace of superconductivity at 4.2 K, as will be discussed later. We have conducted magnetic torque measurements on CaSn$_3$ single crystals in a 31T resistive magnet at the National High Magnetic Field Laboratory (NHMFL) in Tallahassee using a cantilever torque magnetometer. The specific heat of CaSn$_3$ single crystals was measured using the adiabatic thermal relaxation technique in the Physical Property Measurement System (Quantum Design).

**III. RSULTS**

**A. Relativistic fermion behavior in CaSn$_3$**

We have observed strong dHvA oscillations in CaSn$_3$ single crystals in the magnetic torque measurements. In Fig. 1b, we present the dHvA oscillation data for field nearly along the [100] direction (denoted as $B//[100]$). We chose such a field orientation for measurements, because the torque signal vanishes when the field is perfectly aligned normal or parallel to the surface of the cantilever tip where the sample is mounted. The dHvA oscillations start to appear from a field as low as 2T at $T$=1.6 K (see the inset to Fig. 1b), and remain visible at temperatures up to 80K (Fig. 1b). Fig. 1c presents the dHvA oscillation patterns after subtracting the non-oscillating background. Strong quantum oscillations at low fields and high temperatures generally imply high quantum mobility, which is verified by the quantitative analyses of the dHvA data as shown below. Furthermore, given the cubic crystal symmetry of CaSn$_3$, one can expect identical oscillation patterns for $B//[100]$ and $B//[010]$, which is indeed observed in our measurements, as shown in supplementary Fig. S2.

The oscillatory toque shown in Fig. 1c clearly displays multi-frequency components, as revealed by the fast Fourier transformation (FFT) analysis. As shown in Fig. 2a, five major frequencies, *i.e.* $F_\alpha$= 49 T, $F_\beta$ = 59 T, $F_\gamma$ = 347 T, $F_\varepsilon$ = 463 T and $F_\eta$ = 678 T, can be resolved by the FFT analyses, implying a complex Fermi surface for CaSn$_3$. From further analysis of dHvA oscillations, we have found evidence for relativistic fermions in CaSn3. In general, the oscillations of magnetization $\Delta M$ in a 3D topological semimetal can be described by the Lifshitz-Kosevich (LK) formula with a Berry phase being taken into account [36]:

$$\Delta M \propto -B^{\frac{1}{2}} R_T R_D R_S \sin\left[2\pi\left(\frac{F}{B}+\gamma-\delta\right)\right] \qquad (1)$$

where $R_T = \alpha T\mu/B\sinh(\alpha T\mu/B)$, $R_D = \exp(-\alpha T_D\mu/B)$ and $R_S = \cos(\pi g\mu/2)$. Here $\mu = m^*/m_0$ is the ratio of effective cyclotron mass $m^*$ to free electron mass $m_0$. $T_D$ is the Dingle temperature, and $\alpha = (2\pi^2 k_B m_0)/(\hbar e)$. The oscillations are described by the sine term with a phase factor $\gamma - \delta$, where $\gamma = \frac{1}{2} - \frac{\phi_B}{2\pi}$ and $\phi_B$ is the Berry phase. The phase shift $\delta$ is determined by the dimensionality of the Fermi surface and takes a value of 0 for 2D or $\pm 1/8$ for 3D cases.

From the LK formula, the effective mass $m^*$ can be obtained from the fit of the temperature dependence of the oscillation amplitude by the thermal damping factor $R_T$. Because the FFT amplitude is used for the fit, the inverse field $1/B$ in $R_T$ is taken as the average inverse field $1/\bar{B}$, defined as $1/\bar{B} = (1/B_{max} + 1/B_{min})/2$, where $B_{max}$ and $B_{min}$ define the magnetic field range used for the FFT. We have extracted nearly zero effective masses, i.e. $(0.024\pm0.001)m_0$, $(0.022\pm0.001)m_0$, $(0.041\pm0.001)m_0$, and $(0.051\pm0.0008)m_0$ for the $\alpha$, $\beta$, $\gamma$ and $\varepsilon$ bands respectively, which fall into the range of the known topological semimetals [37]. For the $\eta$ band with the highest frequency, a reliable fit is not possible, since the $\eta$-FFT amplitude damps out quickly with increasing temperature and there are only limited data points (see Fig. 2a). Moreover, there are another two additional frequencies near 400T. Since their FFT amplitudes are also quickly damped with increasing temperature, we did not include them in our detailed analyses.

In addition to light effective mass, high quantum mobility and π Berry phase are also important characteristics of relativistic fermions. For the multi-frequency oscillations seen in $CaSn_3$, these parameters cannot be directly obtained through the conventional approaches ( *i.e.* the Dingle plot and the Landau level fan diagram) but can be extracted through a fit of the oscillation pattern to the multiband LK formula, which is generalized from the single band LK model (Eq. 1). This method has been shown to be efficient for the analyses of multi-frequency quantum oscillations in several topological semimetal systems [37–40]. With the effective masses and oscillation frequencies being the known parameters, we can fit the dHvA oscillation patterns at 1.6 K by the multiband LK formula. In order to achieve more accurate fits, we have separated the low frequency oscillation components ($F_α$ and $F_β$) from the high frequency components ($F_γ$ and $F_ε$) and fit them individually. The highest frequency ($F_η$) component, however, has been filtered out since an accurate effective mass cannot be obtained as stated above. As shown in Figs. 2c and 2d, the two-band LK-formula fits both the low- and high-frequency oscillation patterns very well, yielding Dingle temperature $T_D$ of 25, 60, 67, and 75 K for the α, β, γ and ε bands, respectively. From $T_D$, we can further derive quantum relaxation time $τ_q = ℏ/(2πk_BT_D)$ and quantum mobility $μ_q = eτ/m^*$. The obtained values of $μ_q$ are 4278, 1188, 779, and 528 cm$^2$/Vs for α, β, γ and ε bands, respectively. From these multiband LK-fits, we have also determined the phase factor -γ-δ as well as the Berry phase $φ_B$ for each band. Owing to the 3D characteristic of these bands as shown below, the phase shift δ takes value of ±1/8 as mentioned above. With this consideration, the Berry phases derived from our analyses are (1.07± 0.25)π, (0.28± 0.25)π, (−0.58± 0.25)π and (1.58 ± 0.25)π for the α, β, γ and ε bands, respectively (the errors of the fitted Berry phases for α, β, γ and ε bands are all in the range of 0.02-0.04π). The Berry phases for the α and ε bands are close to the ideal value of π. This result, combined with the nearly zero cyclotron mass derived from the temperature dependence of oscillation amplitude, implies relativistic nature of the electrons hosted by these bands.

We have further measured the angular dependences of the quantum oscillations to reveal the Fermi surface morphology of $CaSn_3$. As shown in Fig. 3a, the oscillation patterns display a clear evolution with the change of field orientation angle from the [100] (defined as $\theta = 0°$) to [010] ($\theta = 90°$) direction (see the inset of Fig. 3a for the measurement setup). Fig. 3b summarizes the angular dependence of the major frequencies, obtained from the FFT analyses of the oscillation patterns shown in Fig. 3a. All of the five major frequencies, $F_\alpha$, $F_\beta$, $F_\gamma$, $F_\varepsilon$ and $F_\eta$, have been probed for all field orientations. Additionally, we have probed two frequency components $F_\lambda$ and $F_\kappa$, when the field is not aligned close to the [100] and [010] directions. The angular dependences of all these frequencies display nearly symmetric patterns with respect to [110] ($\theta = 45°$) [*i.e.*, $F(\theta) \approx F(90°-\theta)$], in agreement with the cubic symmetry of $CaSn_3$. These results clearly indicate a 3D Fermi surface for $CaSn_3$.

Although ref. [35] has reported the calculated Fermi surface of $CaSn_3$ and found it is extremely complex and consists of a number of pockets, no quantitative information on the sizes of the pockets is given. Therefore, we are unable to make quantitative comparison between our experimental results and the calculated Fermi surfaces. But, qualitatively, the quantum oscillation frequencies probed in our experiments are consistent with the calculated Fermi surface. Specifically, those lower oscillation frequencies (i.e. $F_\alpha$, $F_\beta$, $F_\kappa$ and $F_\lambda$) are likely to correspond to those small/point-like Fermi surfaces shown in Fig. 4c and 4e in ref. [35], while those higher frequencies (i.e. $F_\gamma$, $F_\varepsilon$ and $F_\eta$) might correspond to those larger Fermi pockets centered around the Brillouin zone center and boundaries (Fig. 4a and 4d in ref.35).

## B. INTRINSIC SUPERCONDUCTIVITY AT 1.178K in CaSn$_3$

In addition to topological fermion properties, CaSn$_3$ also exhibits intrinsic superconductivity below 1.178 K. As shown in Fig. 4, we observed a significant superconducting transition peak near 1.2 K in specific heat. CaSn$_3$ was previously reported to be superconducting with $T_c^{onset} \sim$ 4.2K [35], which appears to contradict our result. However, our detailed studies show that the $T_c^{onset} \sim$ 4.2 K superconductivity, which is also clearly seen in our specific data for one of the measured samples (*i.e.* a small specific heat peak at 3.72 K in sample 1, as denoted an arrow in Fig. 4a), is not an intrinsic property of CaSn$_3$, but results from an impurity phase of Sn present on the sample surface, as has been clearly demonstrated in supplementary material. Such surface Sn impurities should be attributed to the surface decomposition of the air sensitive CaSn$_3$ and is responsible for the superconducting transition near 3.72K in specific heat. When we measured a fresh sample which was exposed to air for a very short period of time (i.e. sample 2 in Fig. 4), we found that the specific heat anomaly near 3.72K due to the superconductivity of Sn on the sample surface became very weak. From the susceptibility data shown in supplementary Fig. S3a, the superconducting volume fraction of Sn is estimated to be ~ 2.4%. Additionally, we also conducted specific heat measurements under a magnetic field of 300 Oe for sample 2 and found its superconductivity at 1.178 K is fully suppressed by this magnetic field, suggesting it its upper critical field is less than 300 Oe.

By subtracting the specific data taken at 300 Oe from the zero-field data, we have obtained the intrinsic superconducting electronic specific heat $C_{es}$ of CaSn$_3$ for sample 2. The obtained data are presented in Fig. 4b, which plots the temperature dependence of $C_e/T - \gamma_n$, where $\gamma_n$ = 3.36 mJ/mol K$^2$ is the Sommerfeld coefficient of CaSn$_3$, obtained from the linear fit shown in the inset of Fig. 4a (note that $C_e/T - \gamma_n$ slightly deviates from zero above $T_c$ (=1.178K) due to the suppression of the superconductivity of impurity Sn on the sample surface by the magnetic field and we have offset it to zero for the BCS fit shown below). From an entropy-conserving construction (see the black dashed line in Fig. 4b), the midpoint transition temperature $T_c^{mid}$ was estimated to be 1.178 K and the specific heat jump $\Delta C/T_c^{mid}$ = 4.03 mJ/mol K$^2$. Using these parameters, $\Delta C/\gamma_n T_c^{mid}$ is estimated to be 1.20, close to the expected value of 1.43 for weak-coupling BCS superconductors, indicating CaSn$_3$ falls into this classification. Furthermore, the superconducting electronic specific heat of CaSn$_3$ can be well-fitted by the single-band isotropic $s$-wave BCS model (see the red fitted curve in Fig, 4b). The reduced gap magnitude $2\Delta/k_B T_c$ obtained from the fit is 3.40, close to the value of 3.53 expected for weak coupling BCS superconductors. The fitted Sommerfeld coefficient $\gamma_n'$ is 3.15 mJ/mol K$^2$. Given that the $\gamma_n$ value obtained from the linear fit in the inset of Fig. 4a is 3.36 mJ/mol K$^2$, the superconducting volume fraction of CaSn$_3$ is estimated to be $\gamma_n'/\gamma_n$ = 94%. Since $\gamma_n$ (=3.36 mJ/mol K$^2$) is slightly overestimated due to the existence of tiny amount of Sn on the sample surface, the actual superconducting volume fraction should be above 94%.

## IV. DISCUSSION

In general, the realization of topological superconductivity requires the presence of spin-polarized topological surface states (TSSs), *i.e.* Dirac-cone type surface states with helical spin polarization. The other requirement is a fully opened bulk superconducting gap in which TSSs appear. When these requirements are met, the topological superconductivity is manifested by a complex surface superconducting order parameter consisting of both spin-singlet and spin-triplet components. The spin-triplet component hosts Majorana fermions. From the above discussions, $CaSn_3$ appears to meet these requirements. As discussed above, our dHvA quantum oscillation studies clearly demonstrate the existence of relativistic fermions in this material, indicating the presence of non-trivial band topology in $CaSn_3$ and possible topological surface states. Moreover, if its topological states are proven to be true, its polarized spins on surface should be aligned along the in-plane direction due to its centrosymmetric crystal structure [32], which should make the surface superconducting order parameter relatively simple. Additionally, the stoichiometric composition of $CaSn_3$ guarantees homogeneous superconductivity, which is a desired condition for probing Majorana surface modes. With these advantages, $CaSn_3$ might be an interesting candidate for a topological superconductor.

Finally, we would like to point out that $CaSn_3$ belongs to a large family of materials with $AuCu_3$-type structure, most of which are superconducting, such as $La_3In$ ($T_c$ =10.4K) [41], $Sr_{1-x}Na_xBi_3$ ($T_c$ = 9K) [42], and $YSn_3$ ($T_c$ = 7K) [43]. The highest $T_c$ among these compounds is about 10K. Since the non-trivial band topology is determined by crystal symmetry, compounds isostructural to CaSn$_3$ may have similar band structure. We expect our work on CaSn$_3$ can inspire band structure studies on other AuCu$_3$-type compounds. If they are also proven to possess non-trivial band topology, they may become candidates of topological superconductors with higher $T_c$.

## VI. CONCLUSION

In summary, from dHvA quantum oscillation studies on CaSn$_3$ single crystal samples, we have found evidence for relativistic fermions in this material. Furthermore, we discovered that this material exhibits intrinsic superconductivity with $T_c \sim 1.178$ K. These findings suggest CaSn$_3$ can be used a platform to explore possible TSC. Further, our finding could also motivate the search for TSC in other members in the large material family with the AuCu$_3$-type structure and superconductivity.

## ACKNOWLEDGEMENTS

This work was supported by the US Department of Energy (DOE) under Grant No. DE-SC0014208. The heat capacity measurements were performed by FNW and PWA with support from the US DOE under BES Grant No. DE-FG02-07ER46420. A portion of this work was performed at the National High Magnetic Field Laboratory, which is supported by National Science Foundation Cooperative Agreement No. DMR-1157490 and the State of Florida.

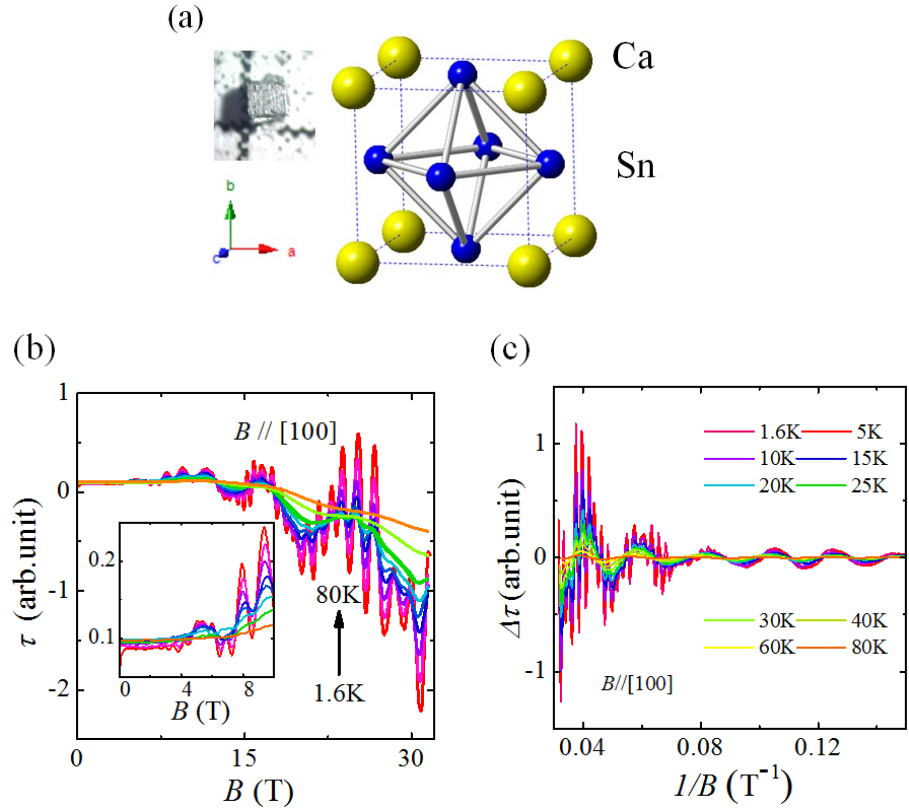

Figure 1: (a) Crystal structure of CaSn$_3$. Inset: an optical image of a CaSn$_3$ single crystal. (b) The field dependence of magnetic torque $\tau$ for CaSn$_3$ at different temperatures from 1.6K to 80K, which shows strong dHvA oscillations. The magnetic field is applied nearly along the [100] direction ($B//[100]$). Inset: enlarged low field oscillations. (c) The oscillatory components $\Delta\tau$ at different temperatures for $B//[100]$.

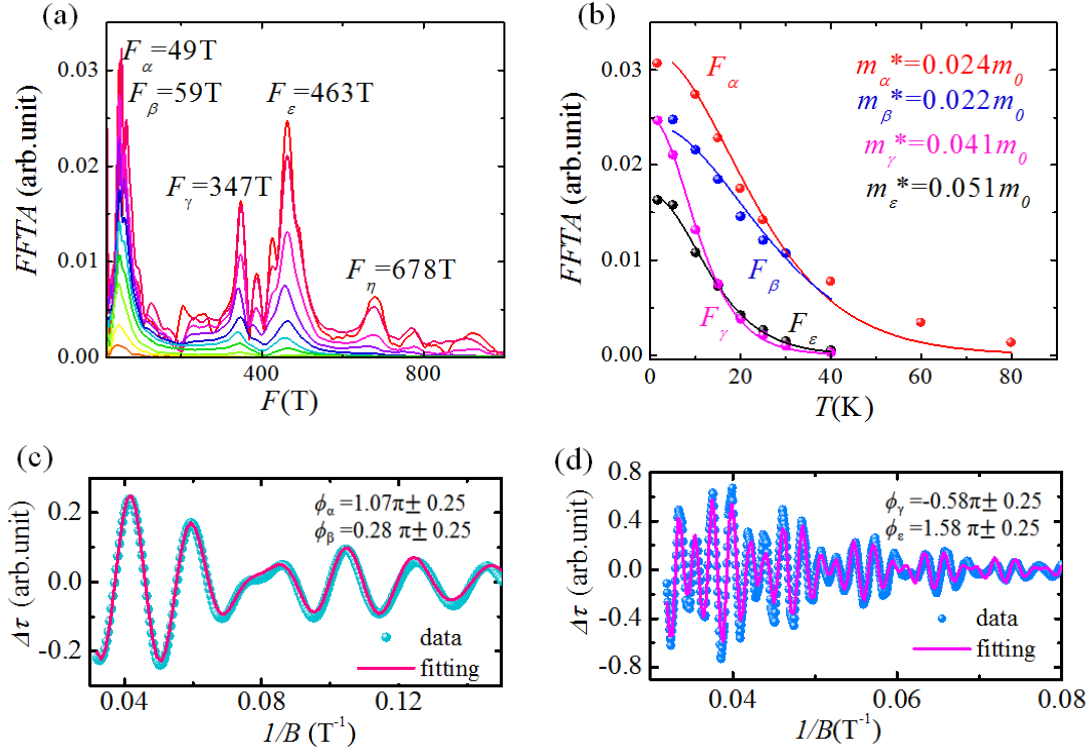

Figure 2: (a) The FFT of the oscillatory magnetic torque $\Delta\tau$ for $B//[100]$. (b) The temperature dependences of the FFT amplitudes for the four major frequencies; the solid lines represent the fits to the LK formula. (c) Low frequency ($F_\alpha$ and $F_\beta$) oscillatory components of magnetic torque obtained after filtering the high-frequency components. (d) High frequency ($F_\gamma$ and $F_\varepsilon$) oscillatory components of magnetic torque obtained after filtering the low-frequency components. The solid curves in (c) and (d) represent the fits of the $T$=1.6K oscillation patterns by the two-band LK formula.

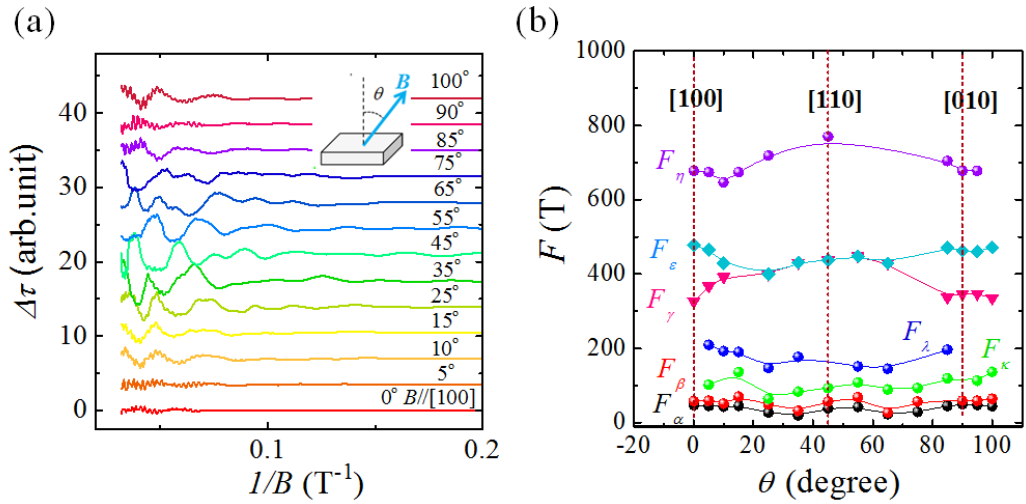

Figure 3: (a) dHvA oscillations of CaSn$_3$ at $T$=1.6K under different magnetic field orientations. Inset: the experimental setup. The data of different $\theta$ have been shifted for clarity. (b) The angular dependence of oscillation frequencies for CaSn$_3$. The vertical dashed lines mark specific crystallographic directions.

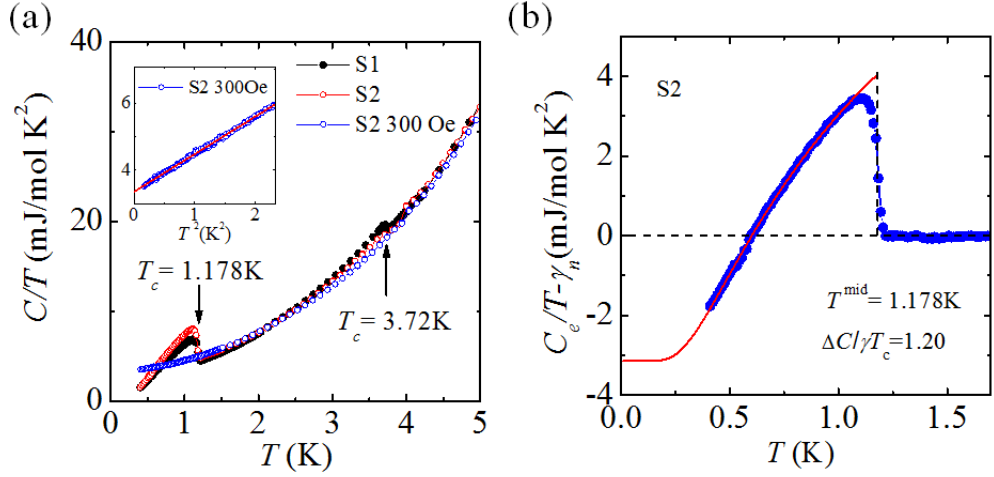

Figure 4: (a) Temperature dependence of specific heat $C(T)/T$ for two CaSn$_3$ single crystal samples. Sample 2 is fresher than sample 1 since it was exposed to air only for a very short period of time. We also measured sample 2 under a magnetic field of 300 Oe (applied along the [100] direction)  Inset: $C(T)/T$ versus $T^2$ at low temperatures for sample 2 measured under the field of 300 Oe. The red line shows the linear fit to $C(T)/T = \gamma_n + \beta T^2$, from which the normal-state Sommerfeld coefficient $\gamma_n$ is estimated to be 3.36 mJ/mol K$^2$. (b) Electronic specific heat $C_e(T)/T$ as a function of temperature after subtracting $\gamma_n$. The red solid curve represents the fit to single-band isotropic $s$-wave BCS model; the entropy balance between superconducting and normal state at $T_c$ is maintained for this fit.

Supplementary Material for

Emergence of intrinsic superconductivity below 1.17 K in the topologically non-trivial semimetal state of CaSn$_3$

1. Powder x-ray diffraction measurements on CaSn$_3$.

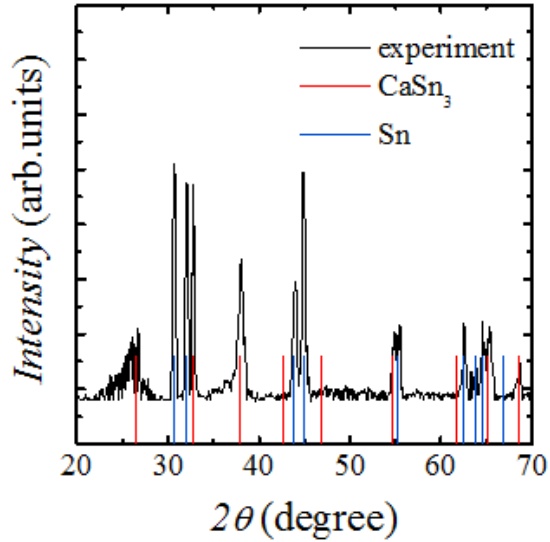

**Supplementary Figure S1**. Powder x-ray diffraction (XRD) pattern recorded on a sample prepared by grinding CaSn$_3$ single crystals to powder in glove box; the sample was protected by Kapton tape during the measurements. Red and blue lines represnet the positions of diffraction peaks of CaSn$_3$ and Sn in the PDF card, respectively. Since Sn is used as flux in crystal growth, the surface of CaSn$_3$ crystals may have tiny amount of residual Sn, although the crystals were separated using centrifuging. We believe the Sn peaks in the XRD spectrum are the result of surface decomposition of CaSn$_3$. This is confirmed by the fact that powder samples prepared in air also exhibits Sn impurity phase as shown in supplementary Figure S4a.

2. dHvA quantum oscillations probed under the field orientations of $B//[010]$

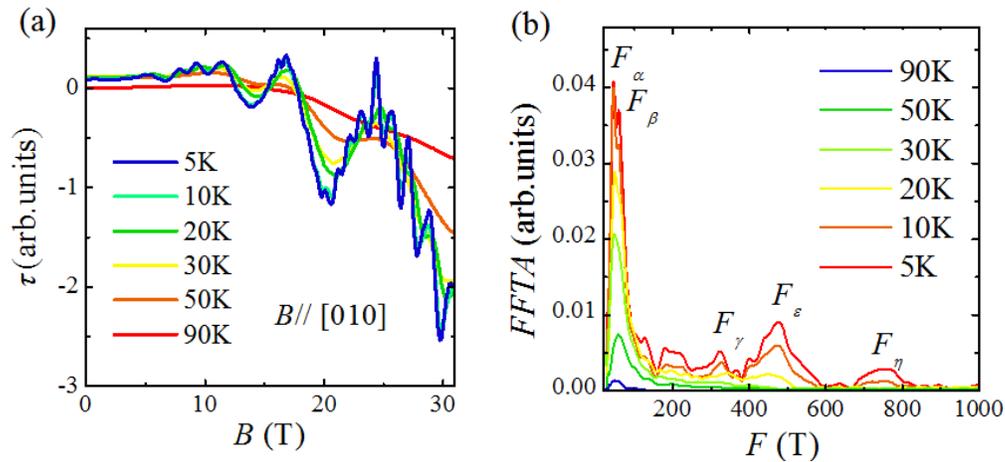

**Supplementary Figure S2**. (a) the field dependence of magnetic torque $\tau$ for CaSn$_3$ at different temperatures from 5K to 90K. The magnetic field is applied nearly along the [010] direction. (b) FFT of the oscillatory magnetic torque, which is obtained by removing the smooth background.

3. Superconductivity of Sn formed from the surface decomposition of CaSn$_3$.

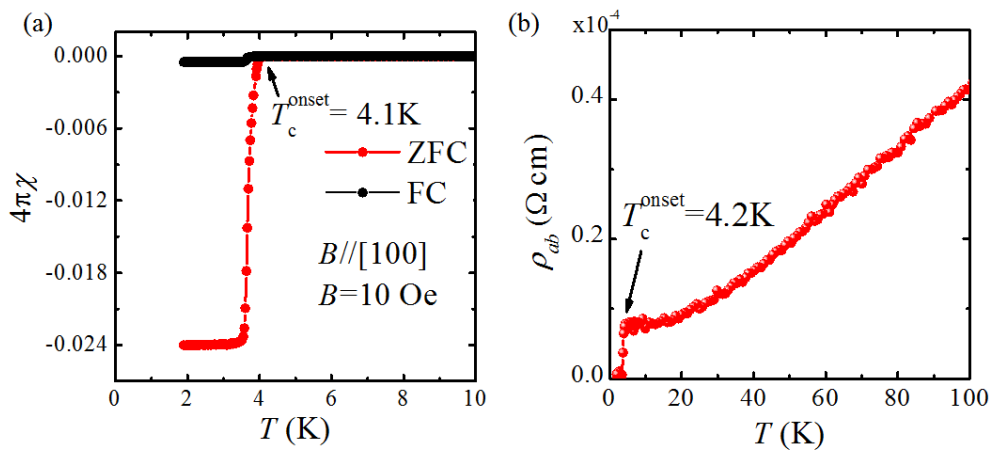

**Supplementary Figure S3**. (a) Temperature dependence of magnetization measured under zero-field-cooling and field-cooling (10 Oe, applied along [100]) histories for a CaSn$_3$ single crystal.

(b) The temperature dependence of resistivity of a CaSn$_3$ single crystal. The superconductivity with $T_c^{onset}$ ~ 4.2 K probed in both susceptibility and resistivity originates from the impurity phase of Sn formed from the degraded surface of CaSn$_3$ single crystal, as discused below.

CaSn$_3$ was previously reported to be superconducting with $T_c^{onset}$ ~ 4.2 K [1]. However, our studies show the $T_c^{onset}$ ~ 4.2 K superconductivity is not an intrinsic property of CaSn$_3$, but originates from an impurity phase of Sn formed from the degraded surface of a CaSn$_3$ single crystal. In fact, we also observed signatures of superconductivity with $T_c^{onset}$ ~ 4.2K in the resistivity and magnetization measurements on our CaSn$_3$ single crystal samples, including zero resistance and diamagnetism as shown in supplementary Figure S3a and S3b, consistent with those previously-reported data in ref. [1]. The superconducting volume fraction of Sn on the sample surface is estimated to be 2.4%. Furthermore, our specific heat data presented in Fig. 4a in the paper shows a small superconducting anomaly peak at 3.72K, which also agrees well the previous observation in specific heat measurements (see Fig. 3 in ref. [1]). Such a weak anomaly peak at 3.72K apparently implies that the superconducting transition at this temperature is non-bulk behavior, which is clearly manifested by the inset to Fig. 4a in the paper where the specific heat divided by temperature $C/T$ is plotted on the scale of $T^2$. The linear dependence of $C/T$ on $T^2$ seen in the inset of Fig. 4a in the temperature range between 3.72 K and 1.2 K (see the red line) indicates significant residual electronic specific heat, with the electronic specific coefficient $\gamma$ = 2.62 mJ/mol K$^2$. In other words, the superconducting transition at 4.2 K happens to only a very small volume fraction, consistent with the ~2.4% superconducting volume fraction of Sn estimated from the susceptibility.

As indicated in the main text, CaSn$_3$ is air sensitive. Its surfaces quickly decompose into some amorphous phases and Sn when the samples are left in air for a few tens of minutes. This is verified by our X-ray diffraction experiments and composition analyses by energy-dispersive spectroscopy (EDS). As shown in supplementary Fig. S1, we observed the phase coexistence of $CaSn_3$ and Sn in the XRD pattern collected on a powder sample prepared by grinding $CaSn_3$ crystals in a glove box and protected by Kapton tape during the measurements; the volume fraction of Sn reflected in the XRD pattern is significant, which was also seen by the authors in ref. [1]. Although we used centrifuging to separate $CaSn_3$ crystals from Sn flux, a small amount of residual Sn on surface is unavoidable. In our EDS measurements, we found the fresh surfaces of $CaSn_3$ crystals indeed have the expected atomic ratio of Ca/Sn = 1/3, while the degraded surfaces have the ratios of 1/4-1/8, indicating the surface decomposition is mostly responsible for the observed rich Sn impurity. We also performed a XRD measurement using a powdered sample prepared by grinding $CaSn_3$ single crystals in air and without Kapton tape-protection during the measurements. The recorded XRD pattern from this experiment (Fig. S4a) does not show any trace of $CaSn_3$, but Sn. Some other amorphous phases must also form from the decomposition though they are not revealed in the XRD pattern. These amorphous phases should originate from oxidation of Ca and Sn during grinding process. After the XRD measurements, we re-collected the powder and put it into a capsule and did magnetization measurements. A superconducting transition at 4.1K, almost identical to the one probed on the $CaSn_3$ single crystal sample was observed in such measurements (see supplementary Figure 4b). The superconducting volume fraction of Sn is about 1.8%, which implies that there exists a small amount of non-oxidized Sn enclosed by Sn oxides. With these results, we conclude that the superconducting transition at 4.1K is most likely due to Sn. We note that the authors of ref. [1] have paid attention to this possibility and made a comparison in the superconducting diamagnetism between their $CaSn_3$ samples and a pure Sn sample. They found their $CaSn_3$ samples exhibit typical type-II superconducting behavior in the diamagnetism, in contrast with the type-I superconducting behavior of pure Sn. Based on this result, they argued that the $T_c$ = 4.2 K superconductivity probed on their sample is not caused by Sn, but the property of CaSn$_3$. In our experiments, we observed similar type-II superconducting behavior in magnetization measurements for the superconducting transition at 4.1 K as well (see Supplementary Figure S4c).

The transition from a type I to type II superconductor for an element superconductor is actually very common when the sample is doped by impurities or structurally disordered. For example, Pb is a type I superconductor, but a small amount of Bi doping (<3%) in Pb can turn it into a type II superconductor [2]. The mechanism of such a transition is actually simple: when impurities are introduced to a type I superconductor, they would shorten the mean free path of electrons and thus reduce the superconducting coherence length $\xi$, which in turn increases the Ginzberg–Landau parameter $\kappa$ ($=\lambda/\xi$, where $\lambda$ is penetration depth). When $\kappa$ is increased above $1/\sqrt{2}$, type I superconductors becomes type II. In our samples, Sn on the surfaces may include impurities such Ca and/or have structural defects, both of which could reduce the mean free path, thus resulting in $\kappa > 1/\sqrt{2}$.

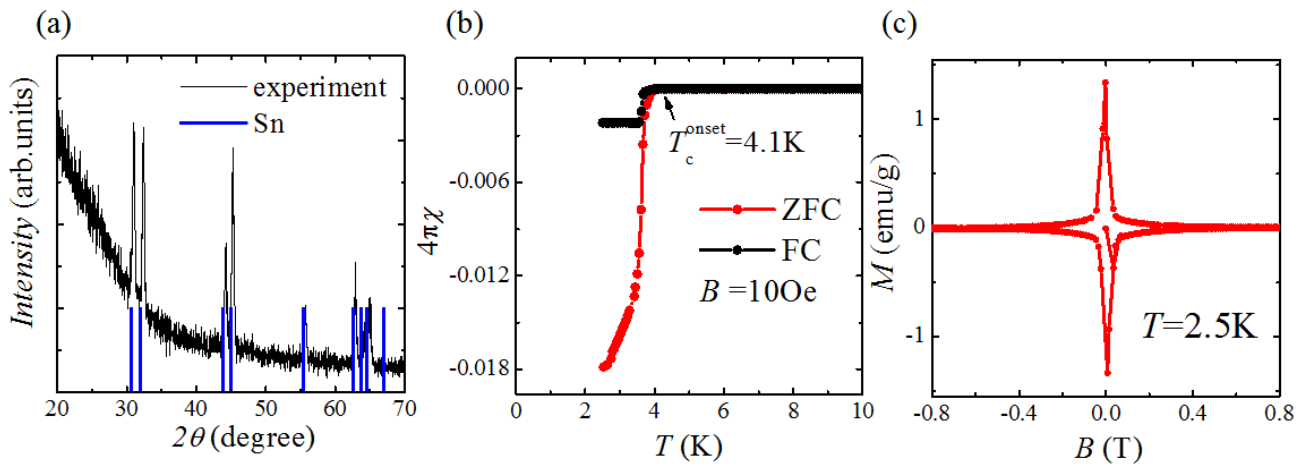

**Supplementary Figure S4**. (a) The x-ray diffraction pattern recoreded on CaSn$_3$ powder prepared by grinding single crystals and exposed in air for several days. The blue lines represnet the positions of Brag diffraction peaks of Sn. The diffraction pattern shows a pure Sn phase, indicating that CaSn$_3$ powder has fully decomposed to Sn and other amorphous phases. (b) Temperature dependence of magnetization measured under zero-field-cooling and field-cooling (10 Oe) conditions for the degraded powder sample used in the XRD measurements. These data show an almost identical superconducting transition at $T_c^{onset}$ =4.1K as the CaSn$_3$ single crystal with degraded suface, indicating that the previously reported superconductivity with $T_c^{onset}$ =4.2K for CaSn$_3$ [1] is due to Sn present on the degraded surface. (c) Isothermal magnetization measured at $T$ = 2.5K for the fully degraded powder sample.